\documentclass[conference]{IEEEtran}
\usepackage{amsmath,citesort,amsfonts,bm} 
\usepackage{verbatim}
\usepackage{color}
\usepackage{graphicx,xspace,color,soul}

\usepackage{epstopdf}
\usepackage{stfloats}

\def\b{{\mathbf b}}
\def\c{{\mathbf c}}

\def\q{{\mathbf q}}

\def\v{{\mathbf v}}
\def\x{{\mathbf x}}
\def\y{{\mathbf y}}
\def\z{{\mathbf z}}
\def\A{{\mathbf A}}
\def\H{{\mathbf H}}

\def\Q{\mathbf{Q}}
\def\L{{\mathbf L}}
\def\V{\mathbf{V}}
\def\U{\mathbf{U}}
\def\cB{{\mathcal B}}

\def\cG{{\mathcal G}}

\def\cV{{\mathcal V}}
\def\bPhi{{\boldsymbol \Phi}}

\def\bEta{{\boldsymbol \eta}}
\def\Lamb{\boldsymbol{\Lambda}}

\begin{document}

\title{Graph-based Joint Signal / Power Restoration for Energy Harvesting Wireless Sensor Networks}


\author{
\IEEEauthorblockN{Megumi Kaneko$^\dagger$, Gene Cheung$^\dagger$, Weng-tai Su$^\star$, and Chia-Wen Lin$^\star$ }
\IEEEauthorblockA{$^\dagger$National Institute of Informatics, Tokyo, Japan 101-8430  \\
$^\star$National Tsing Hua University, Hsinchu, Taiwan\\
Email: \{megkaneko, cheung\}@nii.ac.jp, wengtai2008@hotmail.com, cwlin@ee.nthu.edu.tw}
}


%


\maketitle

\begin{abstract}
---
The design of energy and spectrally efficient Wireless Sensor Networks (WSN) is crucial to support the upcoming expansion of IoT/M2M mobile data traffic.
In this work, we consider an energy harvesting WSN where sensor data are periodically reported to a Fusion Center (FC) by a sparse set of active sensors. Unlike most existing works, the transmit power levels of each sensor are assumed to be unknown at the FC in this distributed setting.
We address the inverse problem of joint signal / power restoration at the FC---a challenging under-determined separation problem.
To regularize the ill-posed problem, we assume both a graph-signal smoothness prior (signal is smooth with respect to a graph modeling spatial correlation among sensors) and a sparsity power prior for the two unknown variables.
We design an efficient algorithm by alternately fixing one variable and solving for the other until convergence.
Specifically, when the signal is fixed, we solve for the power vector using Simplex pivoting in linear programming (LP) to iteratively identify sparse feasible solutions, locally minimizing an objective.
Simulation results show that our proposal can achieve very low reconstruction errors and outperform conventional schemes under reasonable assumptions\footnote{This work is partially supported by the Grants-in-Aid for Scientific Research (Kakenhi) no. 26820143 from the Ministry of Education, Science, Sports, and Culture of Japan.}.

\end{abstract}




%
\IEEEpeerreviewmaketitle

\section{Introduction}
\label{sec:intro}
With 50 billion Internet-of-Things (IoT) devices forecasted for the 2020s, an exponential growth in the amount of mobile data traffic is expected. 
One key issue is the design of wireless access protocols with high spectral and energy efficiency to support Machine-to-Machine (M2M) communications, including control and sensing. In particular, most of the wireless big data traffic generated by M2M applications, such as smart grids, smart homes and intelligent transportation systems, will be supported by Wireless Sensor Networks (WSNs)~\cite{IEC14feb}. A WSN is composed of a large number of low power nodes deployed over an area to detect some physical phenomena such as temperature, humidity or CO2 emissions. Each sensor observes its local environment and transmits its measurement values to a Fusion Center (FC), whose goal is to accurately reconstruct the desired data over the wide area. In addition, wireless sensors are severely battery-limited, so minimizing their power consumption is also important.

To reduce the radio resource and energy consumptions, many works have developed wireless sensing methods based on Compressed Sensing (CS) theory~\cite{Hau08mar,Faz11feb,Yan13sep}. Since the monitored data are typically correlated in space, time and/or frequency, they can be well represented by a sparse vector in a transform domain. CS theory stipulates that the sparse vector in a transformation basis can be recovered with high accuracy from a few random projections onto another incoherent basis. As one representative example, \cite{Yan13sep} proposed a wireless CS-based reconstruction method Energy Harvesting (EH) WSNs that exploits the sparse nature of the monitored environment while guaranteeing stable data recovery even with heterogeneous Signal-to-Noise Ratios (SNR) due to disparate distances between sensors and the FC. 

Recently, leveraging on recent advances in graph signal processing (GSP) \cite{Shu13may}, a graph-based sparse signal recovery technique was proposed in~\cite{Zhu12mar} to estimate signals that are smooth with respect to a carefully constructed graph. In WSNs, such a graph would model correlations among data samples collected by spatially distributed sensors. It is shown that if the original signal is smooth with respect to the network topology, the signal is linearly compressible through the Graph Fourier Transform (GFT). Similar in principle to CS, this approach enables a good signal reconstruction by only activating $K \ll N$ sensor nodes, thereby saving radio and energy resources. 
However, in most previous works such as \cite{Hau08mar,Faz11feb,Yan13sep,Zhu12mar}, transmit power allocation at each sensor node has not been considered. For example, in \cite{Yan13sep}, each sensor makes use of the harvested power for data transmission, but the transmit power is integrated in the measurement matrix, which is assumed to be perfectly known at the FC. This is hardly realistic in the distributed setting, where typically each sensor sets its own transmit power in a local manner.

In this work, we consider a wireless powered communication network for energy harvesting sensor nodes with energy transfer in the downlink and data transmission in the uplink, as done in \cite{Yan13sep,Ju14jan}. Unlike \cite{Yan13sep,Zhu12mar}, the transmit powers are set independently at individual sensors and hence are unknown at the FC. This is consistent with the distributed setting of WSNs, in particular with EH as the amount of allocated power by each sensor depends on the amount of energy harvested locally. This gives rise to an intricate inverse problem for both the transmit power vector and the signal vector, whose coefficients are multiplied in the received signal at FC. 

Our main contributions are as follows:

\noindent 1) We formulate a joint signal/power restoration problem given only sparse observations in a WSN, where multiple sensor nodes send their local measurements to the FC. To regularize the ill-posed problem, for signal prior, we assume that the signal is smooth with respect to a graph modeling correlations among nodes, leveraging recent graph-signal restoration works \cite{pang17}.
For power prior, we assume that transmit power vector is $K$-sparse with known lower and upper bounds. 

\noindent 2)  We show that the joint problem can be solved alternately by solving one variable at a time while holding the other fixed until convergence.
Specifically, we solve the power restoration sub-problem via a novel implementation of \textit{Simplex pivoting} in linear programming (LP) to identify consecutive sparse feasible solutions \cite{papadimitriou98} to minimize an objective.
\textit{To the best of our knowledge, we are the first in the literature to tackle this inverse separation problem, where signal and power are sought simultaneously from sparse multiplied observations in wireless systems.}


\noindent 3) Finally, simulation results show that our proposed method outperforms conventional reconstruction schemes based on CS. By reducing the number of required observations and of active sensors for a given reconstruction level, our proposed scheme reduces the radio resource and energy consumption, which are crucial issues for battery-limited WSNs.

\vspace*{-0.2cm}
\section{System Model}
\label{sec:system}

\begin{figure}[t]
\centering
\includegraphics[scale=0.25]{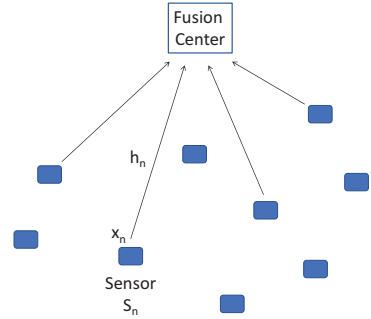}
\vspace*{-0.7cm}
\caption{System Model: Wireless Sensor Network}
\label{fig:WSN}
\end{figure}

We consider the WSN in Fig.\;\ref{fig:WSN} composed of $N$ wireless sensors with EH capabilities and one FC.
Each sensor $S_n, n=1 ,\ldots, N$, locally monitors the environment and sends its measurement $x_n \in \mathbb{R}$ to the FC. These measurements are represented by vector $\mathbf{x}$ of length $N$, which we assume to be \textit{smooth}; \textit{i.e.}, physical observations, such as temperatures, humidity and CO2 levels, are typically spatially correlated. 
Each sensor transmits its scalar data $x_n$ during $M<N$ time slots to reduce the radio resource and energy consumptions. 
Similar to \cite{Hau08mar,Yan13sep,Faz11feb}, each sensor multiplies its measurement data by a pseudo-random signature sequences (\textit{e.g.}, following Bernoulli distribution) $\bm{\phi}_n=[\phi_{1,n}, \cdots, \phi_{M,n}]^{T}$ of size $M\times 1$ $\in$ $\{0,1\}^N$. These vectors $\bm{\phi}_n$ are concatenated to form the columns of measurement matrix $\bPhi$ of size $M\times N$. Since $\bPhi$ is composed of the signature sequences (identification vectors) of all sensors, it is known at the FC as in~\cite{Yan13sep,Faz11feb}. 
Note that unlike~\cite{Ju14jan} that required a centralized scheduler to guarantee that only one sensor measurement is received per time slot, here each sensor transmits in a distributed manner, so that a linear combination of different sensor measurements is received at any time slot $i=1, \ldots, M$. 

Each sensor transmits with power $p_n$ during the $M$ slots constrained by the available power in that time frame. This power budget depends on the amount of harvested energy by Wireless Power Transfer (WPT) during the previous downlink frame as in~\cite{Ju14jan}. Without loss of generality, each $x_n$ is assumed to be a zero-mean random variable following a distribution to be specified later and with normalized power $\mathbb{E}[|x_n|^2]=1$. Defining the transmit power (energy) vector $\bm{\eta}=[\sqrt{p_1}, \cdots, \sqrt{p_N}]^T$ and denoting $\mathrm{diag}(\bm{\eta})$ the diagonal matrix with elements $(\eta_1, \cdots, \eta_N)$ on its diagonal, the received signal at the FC is given by vector $\mathbf{y} \in \mathbb{C}^{M\times 1}$ 
\begin{equation}
	\mathbf{y}=\bPhi \H \mathrm{diag}(\bm{\eta}) \mathbf{x} + \mathbf{w}
\label{eq:y}	
\end{equation}
for each frame, where $\H=\mathrm{diag}(h_1, ..., h_N) \in \mathbb{C}^{N \times N}$ is the diagonal channel matrix gathering the channel coefficients between each sensor $S_n$ and the FC, as each sensor is assumed to transmit in a block Rayleigh fading channel~\cite{Ju14jan}. 
$\mathbf{w}$ is the Additive White Gaussian Noise (AWGN) vector.\footnote{As in many works such as~\cite{Yan13sep}\cite{Zhu12mar}, $\mathbf{x}$ is assumed to take continuous values as, information theoretically, such analog sensing enables lower reconstructed signal distortion compared to digital~\cite{Gas06jul}. However, practical issues with discrete modulation/coding will be considered in the follow-up work.}
Next, we overview the basics of GSP used in our proposal.

\section{Basics of Graph Signal Processing}
\label{sec:gsp}
\vspace{-0.05in}
\subsection{Graph Definition}
\label{subsec:defn}

\vspace{-0.05in}
A graph $\mathcal{G}(\cV, \mathcal{E}, \mathbf{W})$ has a set $\cV$ of $N$ nodes and a set $\mathcal{E}$ of $Q$ edges.
Each edge $(n,k) \in \mathcal{E}$ connecting nodes $n$ and $k$ is undirected with positive weight $w_{n,k}$. 
A graph-signal $\mathbf{x} \in \mathbb{R}^N$ on $\mathcal{G}$ is a discrete signal of dimension $N$---one label $x_n$ for each sensor node $S_n$ in $\mathcal{V}$.


\vspace{-0.05in}
\subsection{Graph Spectrum}
\label{subsec:spectrum}

\vspace{-0.05in}
Given edge weight (adjacency) matrix $\mathbf{W}$, we define a diagonal \textit{degree matrix} $\mathbf{D}$, where $d_{n,n} = \sum_{k} w_{n,k}$.
A \textit{combinatorial graph Laplacian matrix} $\mathbf{L}$ is $\mathbf{L} = \mathbf{D} - \mathbf{W}$~\cite{Shu13may} . %
As $\mathbf{L}$ is symmetric, it can be eigen-decomposed into $\L = \V \Lamb \V^T$ (Spectral Theorem),
where $\Lamb$ is a diagonal matrix containing real eigenvalues $\lambda_j$, and $\V$ is an eigen-matrix composed of orthogonal eigenvectors $\v_j$ as columns.
If $w_{n,k}$ are non-negative, then $\mathbf{L}$ must be \textit{positive semi-definite} (PSD), meaning that $\lambda_j \geq 0, \forall j$ and $\x^T \L \x \geq 0, ~\forall \x$.
Non-negative eigenvalues $\lambda_j$ can be interpreted as \textit{graph frequencies}, and eigenvectors $\v_j$ interpreted as corresponding graph frequency components. Together they define the \textit{graph spectrum} for graph $\cG$.


\vspace{-0.05in}
\subsection{Graph-Signal Smoothness Prior}
\label{subsec:smooth}

\vspace{-0.05in}
For graph $\cG$ with positive edge weights, signal $\x$ is considered \textit{smooth} if each label $x_n$ on sensor node $S_n$ is similar to labels $x_k$ on neighboring nodes $k$ with large $w_{n,k}$.
In the graph frequency domain, it means that $\mathbf{x}$ contains mostly low graph frequency components; \textit{i.e.}, coefficients $\boldsymbol{\alpha} = \V^T \x$ are zeros or very small for high frequencies.
The smoothest signal is the constant vector $\mathbf{1}$---the first eigenvector $\v_1$ for $\L$ corresponding to the smallest eigenvalue $\lambda_1 = 0$.

Mathematically, we can write that a signal $\mathbf{x}$ is smooth if its \textit{graph Laplacian regularizer} $\x^T \L \x$ is small \cite{pang17}. 
Graph Laplacian regularizer can be expressed as:
\begin{align}
\x^T \L \x  & = \sum_{(n,k) \in \mathcal{E}} w_{n,k} \left( x_n- x_k \right)^2 = \sum_{j} \lambda_j \, \alpha_j^2
\label{eq:smoothness}
\end{align}
Because $\L$ is PSD, $\x^T \L \x$ is lower-bounded by 0.
We will use $\x^T \L \x$ as the smoothness prior in the sequel.

\section{Proposed Method}
\label{sec:proposed}
\subsection{Protocol Description}

We first describe our proposed protocol for signal transmission and reconstruction, composed of two phases.

\noindent{\underline{Phase 1}} : 
As in the harvest-and-transmit protocol of~\cite{Ju14jan}, the Phase 1 consists of the downlink transmission where 
the FC (Hybrid-Access Point~\cite{Ju14jan}) transmits energy to the sensors by WPT. Each sensor harvests
energy $\xi_n$ during time $T_e$,
\begin{equation}
	\xi_n=\rho_n |g_n|^2 P_{e} T_e,
	\label{eq:xi}
\end{equation}
where $\rho_n \in (0,1]$ is the energy efficiency of sensor $S_n$, $g_n$ the channel gain between the FC and sensor $S_n$ and $P_{e}$ the transmit power of the FC.

\noindent{\underline{Phase 2}} : 
In the second phase, each sensor $S_n$ sends its measured data $x_n$ to the FC during a frame of $M$ time slots according to the model in Section \ref{sec:system}. Being severely battery-limited, each sensor will be active with a given transmission probability $\psi_n$, known at the FC as in~\cite{Yan13sep}.   
Denote by $K$ the average number of active sensors per frame. 
Our goal is to provide a good reconstruction performance for $\mathbf{x}$ even when $K \ll N$, \textit{i.e.}, when the set of active sensors is sparse compared to the total number of sensors~\cite{Hau08mar,Faz11feb}.   
For simplicity, we assume as in \cite{Ju14jan} that each active sensor transmits with power $p_n \leq \min\{P_{\mathrm{max}},\xi_n\}$, where $P_{\mathrm{max}}$ is the maximum allowed transmit power per sensor and $\xi_n$ is the energy harvested during the first phase. 
However, exactly which sensors are active and what is the transmit power level of each sensor $p_n$ for each frame are unknown at the FC under this distributed setting. The received signal at the FC is given by (\ref{eq:y}).

\subsection{Problem Formulation}

The objective of our method is to jointly optimize over the transmit power vector $\bm{\eta}$ and the reconstructed signal $\mathbf{\hat{x}}$ such that the reconstruction error is minimized. 
In this problem, we focus on smooth data signals $\mathbf{x}$, \textit{i.e.}, spatially correlated signals among the different sensor readings.
Hence, the considered optimization problem can be formulated as a weighted sum among three terms: i) the reconstruction error (fidelity term), ii) the graph-signal smoothness prior defined in Section \ref{subsec:smooth} with parameter $\mu$, and iii) sum of the negative log probabilities of active sensors with parameter $\gamma$,

\begin{footnotesize}
\begin{eqnarray}
   &  \underset{\bm{\eta},\mathbf{x}}{\mathrm{min}} \ J(\bm{\eta},\mathbf{x})=||\mathbf{y}-\bPhi \H \mathrm{diag}(\bm{\eta}) \mathbf{x}||_2^2 + \mu \; \x^T \L \x 
   - \gamma \underset{i | \eta_i > 0}{\sum} \log \psi_i \nonumber\\
   & \text{s.t.} \quad \mathbf{0} \leq \bm{\eta} \leq \bm{\eta}_{\mathrm{max}}, 
   ~ \| \bEta \|_0 = K
\label{eq:opt}
\end{eqnarray}
\end{footnotesize}\noindent
where the constraints are on the maximum transmit power constraints at each sensor $\bm{\eta}_{\mathrm{max}}=\eta_{\mathrm{max}} \mathbf{1}$ with $\eta_{\mathrm{max}} = \sqrt{P_{\mathrm{max}}}$ ($\mathbf{1}$ is the vector of ones of size $N$) and their non-negativity, and a $K$-sparse constraint (given on average $K$ sensors are expected to be active). 




\vspace*{-0.2cm}
\section{Optimization}
\label{sec:opt}
The problem formulated in (\ref{eq:opt}) is a non-convex optimization problem with respect to variables $\bEta$ and $\x$, making it difficult to solve.
Thus we employ an alternating search strategy where we fix one variable $\bEta$ and optimize the other $\x$ and vice versa until convergence.
We summarize our algorithm below.


\vspace{0.1in}
\noindent{\underline{\textbf{Step 0}: initialize $\mathbf{x}$}}
\vspace{0.05in}

Before we start our alternating algorithm, we first initialize $\hat{\x}_o$ as follows. 
Because $\x$ is likely a smooth signal, as a first-order approximation, we can assume $\hat{\x}_o = c \mathbf{1}$, for some constant $c$. 
To find the optimal $c$, we write:
\begin{equation}
\min_{c} \| \y - \bPhi \H \mathrm{diag}(\hat{\bEta}_o) c \mathbf{1} \|_2^2
\label{eq:optC}
\end{equation}
where we assume $\hat{\bEta}_o = \bEta_{\max}$.
We take the derivative of (\ref{eq:optC}) with respect to $c$ and set it to zero to derive the optimal $c^*$:
\begin{equation}
c^* = \frac{\y^T \bPhi \H \mathrm{diag}(\hat{\bEta}_o) \mathbf{1}}{\mathbf{1}^T (\bPhi \H \mathrm{diag}(\hat{\bEta}_o))^T \bPhi \H \mathrm{diag}(\hat{\bEta}_o) \mathbf{1}}
\label{eq:optC2}
\end{equation}
Because (\ref{eq:opt}) is non-convex, the alternating algorithm converges in general to a local minimum, and the initialization is very important.
Empirically we found initializing $\hat{\x}_o = c \mathbf{1}$ where $c$ is computed using (\ref{eq:optC2}) to be very effective. 

\vspace{0.1in}
\noindent{\underline{\textbf{Step 1}: solve $\bm{\eta}$ for fixed $\mathbf{x}$}}
\vspace{0.05in}

Given fixed $\x = \hat{\x}_{i-1}$, we solve for the optimal $\bEta$.
Observing that $\mathrm{diag}(\bEta) \x = \mathrm{diag}(\x) \bEta$, (\ref{eq:opt}) now simplifies to: 
\begin{eqnarray}
   &  \underset{\bEta}{\mathrm{min}} \ J(\bEta)= \| \mathbf{y}-\bPhi \H \mathrm{diag}(\hat{\x}_{i-1}) \bEta \|_2^2 - \gamma \underset{i | \eta_i > 0}{\sum} \log \psi_i \nonumber\\
   & \text{s.t.} \quad \mathbf{0} \leq \bEta \leq \bEta_{\max}, ~ \| \bEta \|_0 = K 
\label{eq:opt_eta}
\end{eqnarray}
While the first term in the objective is quadratic, the second term tabulates the probabilities of the non-zero entries in $\bEta$, which, together with the $K$-sparse constraint, makes the optimization difficult to solve. 
We focus our discussion on solving (\ref{eq:opt_eta}) in the next subsections.

\vspace{0.1in}
\noindent{\underline{\textbf{Step 2}: solve $\mathbf{x}$ for fixed $\bm{\eta}$}}
\vspace{0.05in}

Given fixed $\bEta = \hat{\bEta}_i$, we solve for the optimal $\x$, where (\ref{eq:opt}) simplifies to:
\begin{eqnarray}
   &  \underset{\mathbf{x}}{\mathrm{min}} \ J(\mathbf{x}) = \| \mathbf{y}-\bPhi \H \mathrm{diag}(\hat{\bEta}_{i}) \mathbf{x} \|_2^2 + \mu \, \x^T \L \x
\label{eq:optx}
\end{eqnarray}

(\ref{eq:optx}) is an unconstrained optimization with two quadratic terms as objective, so we take the derivative with respect to  $\x$ and equate it to $0$ to solve for the optimal $\hat{\x}_i$ at iteration $i$:
\begin{align}
\hat{\x}_i =& \left\{(\bPhi \H \mathrm{diag}(\hat{\bEta}_{i}))^H (\bPhi \H \mathrm{diag}(\bm{\hat{\eta}}_{i}))+\mu \L \right\}^{-1} \nonumber \\
& \times (\bPhi \H \mathrm{diag}(\hat{\bEta}_{i}))^H \mathbf{y},
\label{eq:optx2}
\end{align}
One can show that the matrix in the first line in (\ref{eq:optx2}) is full rank and invertible.

Steps 1 and 2 will be executed alternately until covergence in $\x$, or a maximum number of iterations is reached.

\subsection{Finding the Optimal Power Vector}
\label{subsec:second}

For simplicity of notation, denote by $\Q = \bPhi \H \mathrm{diag}(\hat{\x}_{i-1})$.  
We solve (\ref{eq:opt_eta}) in two parts as follows.  
We first ignore the probability term in the objective and the $K$-sparse constraint, resulting in the following convex optimization problem:
\begin{eqnarray}
   &  \bEta^* = \underset{\bEta}{\mathrm{argmin}} \| \mathbf{y}-\Q \bEta \|_2^2 \quad \text{s.t.} \quad \mathbf{0} \leq \bEta \leq \bEta_{\max}.
\label{eq:opt_eta2}
\end{eqnarray}
Solving (\ref{eq:opt_eta2}) means that within a spherical solution space centered at $\y$ and radius $r = \| \y - \Q \bEta^* \|_2^2$, there exists a feasible solution $\bEta^*$ that satisfying the power constraint. 

Given this solution $\bEta^*$, in the second part we first define a set of \textit{linear} constraints for each $i$ of $M$ dimensions of $\y$:
\begin{align}
y_i - \epsilon_i \leq [\Q]_i \bEta \leq y_i + \epsilon_i, 
~~~ \epsilon_i = | y_i-[\Q]_i \bEta^* |
\label{eq:opteps}
\end{align}
where $[\Q]_i$ denotes the $i$-th row of matrix $\Q$. 
Geometrically, these $M$ linear constraints define a tighter $M$-dimensional polytope inside the earlier spherical solution space. 
Note that solution $\bEta^*$ in the first part is one feasible solution for these linear constraints; the constraints ensure that the obtained solution in the second part has fidelity error no worse than $r$, since the new search space (polytope) is inside the previous search space (sphere).

We can now reformulate (\ref{eq:opt_eta}) as follows:
\begin{align}
\min_{\bEta} \underset{i | \eta_i > 0}{\sum} - \log \psi_i ~~~ \mbox{s.t.} ~~
\left\{ \begin{array}{l}
\| \bEta \|_0 = K \\
\y - \bm{\epsilon} \leq \Q \bEta \leq \y + \bm{\epsilon} \\
\mathbf{0} \leq \bEta \leq \bEta_{\max}
\end{array} \right.
\label{eq:obj2}
\end{align}
where the quadratic fidelity term is replaced by the $M$ linear constraints. 

Unlike typical sparse coding problems \cite{elad06}, (\ref{eq:obj2}) is more difficult due to the additional linear constraints.
\textit{Our key idea is to employ ``pivoting" in Simplex linear programming (LP) algorithm \cite{papadimitriou98} to identify these feasible solutions to greedily optimize (\ref{eq:obj2}).}
Geometrically, the feasible space in a LP formulation is the intersection of defined half-spaces---a polytope. 
Because it is known that an optimal LP solution (if LP is feasible and bounded) exists at a vertex of a polytope, Simplex marches from one polytope vertex to a neigbhoring one (called \textit{pivoting}) in search for an optimal solution \cite{papadimitriou98}.
Though our objective in (\ref{eq:obj2}) is not linear in $\bEta$, one can nonetheless show easily that an optimal solution to (\ref{eq:obj2}) exists at a polytope vertex (if (\ref{eq:obj2}) is feasible and bounded).
Thus we can also employ Simplex pivoting to march among neighboring polytope vertices to locally optimize (\ref{eq:obj2}).                   

We first overview LP, then describe pivoting operation in Simplex. 
Finally, we describe how we employ Simplex pivoting to optimize (\ref{eq:obj2}).

\subsection{Linear Programming in Standard Form}
\label{subsec:LP}

Linear programming (LP) seeks an optimal $N$-dimensional solution vector $\z \in \mathbb{R}^N$ that minimizes a linear objective subject to non-negativity and $M$ additional linear constraints. 
The \textit{standard form} of LP can be written as \cite{papadimitriou98}:
\begin{equation}
\min_{\z} \c^T \z
~~~ \mbox{s.t.} ~ 
\left\{ \begin{array}{l}
\A \z = \b \\
\z \geq \mathbf{0}
\end{array} \right.
\label{eq:LP}
\end{equation}
where $\c \in \mathbb{R}^N$ is the cost vector, $\A \in \mathbb{R}^{M \times N}$ and $\b \in \mathbb{R}^M$ specify the $M$ linear constraints, and $\z \geq \mathbf{0}$ means each entry in $\z$ must be non-negative, \textit{i.e.} $z_i \geq 0, \forall i \in \{1, \ldots, N\}$. 
LP can be solved in polynomial time using a number of known algorithms such as interior point and ellipsoid methods \cite{papadimitriou98}. 
In this paper, we focus on the \textit{Simplex} method. 
The reason is because Simplex marches from one sparse feasible solution $\z_t$ at one polytope vertex to another $\z_{t+1}$ at another vertex via pivoting as the objective is iteratively minimized.
Given that (\ref{eq:obj2}) contains both linear and sparsity constraints, Simplex pivoting enables us to identify these feasible sparse solutions as we iteratively minimize the objective in (\ref{eq:obj2}).


\subsection{Definitions in Simplex}

To facilitate discussion, we first define important concepts in Simplex. 
A \textit{basis} $\cB$ is a particular selection of $M$ of $N$ columns in $\A$, $\cB(1), \ldots, \cB(M)$, such that the $M \times M$ square matrix $\A_{\cB} = [A_{\cB(1)} \cdots A_{\cB(M)}]$ is full rank.
Corresponding to $\cB$ is a reordered \textit{basic solution} $[\z_{\cB} ~ \z_{\bar{\cB}}]^T$, where $\z_{\cB}$ collects entries $i$ of solution $\z$ computed as:
\begin{align}
z_i = \left[ \A_{\cB}^{-1} \b \right]_j ~~ \mbox{if} ~ i = \cB(j) 
\end{align}
where $[\v]_j$ denotes the $j$-th entry of vector $\v$. 
$\z_{\bar{\cB}}$ is simply a length $N-M$ zero vector. 
Note that by definition of basis $\cB$, $\A_{\cB}$ is full rank and invertible. 
Note further that basic solution $\z_{\cB}$ has $N-M$ entries equal to $0$, or $M$-sparse\footnote{A \textit{degenerate} basic solution can have more than $(N-M)$ zeros in $\z$ \cite{papadimitriou98}. Though they occur rarely in practice, our theory can be easily extended to account for these cases as well but is left for future work.}. 

A basic solution $\z = [\z_{\cB} ~ \z_{\bar{\cB}}]^T$ clearly satisfies $\A \z = \b$:
\begin{align}
\A \z = [\A_{\cB} ~ \A_{\bar{\cB}}] 
\left[ \begin{array}{c}
\z_{\cB} \\
\z_{\bar{\cB}}
\end{array} \right] = \A_{\cB} \z_{\cB} + \mathbf{0} = \b
\end{align}
where we rearrange the columns of $\A$ so that the first $M$ columns are the basis $\cB$.
A \textit{basic feasible solution} (bfs) $[\z_{\cB} ~ \bar{\z_{\cB}}]^T$ in addition satisfies the non-negativity constraint, \textit{i.e.}, $\z_{\cB} \geq \mathbf{0}$.

\subsection{Pivoting in Simplex}
\label{subsec:pivot}

Pivoting is the process of moving from one bfs to another, by pushing in one non-basic column in $\A$ into basis $\cB$, replacing one current basic column. 
To understand the underlying mechanism, we consider the following.
Let $\z = [\z_{\cB} ~ \z_{\bar{\cB}}]^T$ be a bfs corresponding to basis $\cB$, which means:
\begin{align}
\sum_{i=1}^M z_i A_{\cB(i)} = \b
\label{eq:lin1}
\end{align}
Because $\A_{\cB}$ has rank $M$, a non-basic column $A_j$, $j \not\in \cB$, can be written as a linear combination of columns in $\A_{\cB}$:
\begin{align}
\sum_{i=1}^M h_{j,i} A_{\cB(i)} = A_j
\label{eq:lin2}
\end{align}
where $h_{j,i}$ denotes the weight coefficient for column $A_{\cB(i)}$ in the linear combination to compose $A_j$. 

If we now linearly combine the two equations as $(\ref{eq:lin1}) - \theta \, (\ref{eq:lin2})$ for a small $\theta$, $\theta > 0$ and rearrange terms, we get:
\begin{align}
\sum_{i=1}^M (z_i - \theta h_{j,i}) A_{\cB(i)} + \theta A_j = \b
\label{eq:lin3}
\end{align}
(\ref{eq:lin3}) states that $\b$ is now a linear combination of $M+1$ columns in $\A$. 
As $\theta$ increases, a coefficient $z_i - \theta h_{j,i}$ may decrease and eventually become zero. 
Because the starting solution is a bfs, $z_i > 0$, and the smallest value $\theta$ at which a coefficient becomes zero is:
\begin{align}
\theta^* = \min_{i \,|\, h_{j,i} > 0} \frac{z_i}{h_{j,i}}
\label{eq:pivot}
\end{align}
Denote by $k$ the index of this entry. 
Mathematically, (\ref{eq:pivot}) means that at $\theta^*$, $\b$ is now a linear combination of a \textit{new} basis $\cB'$ of $M$ columns in $\A$, where previously non-basic column $j$ replaced column $k$. 

Practically, this means that pushing a non-basic column $j$ in $\A$ into basis $\cB$ will replace a column with index computed using (\ref{eq:pivot}). 
At any given iteration, a user has the freedom to choose one from $N-M$ non-basic columns in $\A$ to enter basis $\cB$. 
We exploit this freedom to optimize (\ref{eq:obj2}) next.


\subsection{LP for Power Vector Restoration}

To use pivoting to select $\bm{\eta}$ in (\ref{eq:obj2}), we first write the constraints in (\ref{eq:obj2}) in standard form (\ref{eq:LP}) as follows:

\begin{small}
\begin{align}
\underbrace{\left[ \begin{array}{c}
\y + \bm{\epsilon} \\
\y - \bm{\epsilon} \\
\bm{\eta}_{\max}
\end{array} \right]}_{\b}
= 
\underbrace{\left[ \begin{array}{cccc}
\Q & \mathbf{1} & \mathbf{0} & \mathbf{0} \\
\Q & \mathbf{0} & -\mathbf{1} & \mathbf{0} \\
\mathbf{1} & \mathbf{0} & \mathbf{0} & \mathbf{1} 
\end{array} \right]}_{\A}
\underbrace{\left[ \begin{array}{c}
\bm{\eta}\\
\q_1 \\
\q_2 \\
\q_3
\end{array} \right]}_{\z}, ~
\left[ \begin{array}{c}
\bm{\eta} \\
\q_1 \\
\q_2 \\
\q_3
\end{array} \right] \geq \mathbf{0}
\label{eq:LPconstraints}
\end{align}
\end{small}\noindent
where we introduce \textit{slack variables} $\q_1$, $\q_2$ and $\q_3$, where $\q_i \geq \mathbf{0}$, so that inequalities can be expressed as equalities in LP standard form.
$\bEta \in \mathbb{R}^N$ and $\y \in \mathbb{R}^M$, and hence the matrix in (\ref{eq:LPconstraints}) has dimension $(2M+N) \times 2(M+N)$. 
Having done this, we employ pivoting to march through bfs to optimize the objective in (\ref{eq:obj2}) as follows.
We first compute an initial bfs using \texttt{linprog} function in Matlab with constraints in (\ref{eq:LPconstraints}) and a cost vector $\c = [-\mathbf{1} ~ \mathbf{0} ~ \mathbf{0} ~ \mathbf{0}]$, so that $\bEta$ is maximized.
Depending on the number of basic columns in basis $\cB$ corresponding to $\bEta$ (number of active sensors, denoted by $g$) at each iteration, we do the following: 
\begin{enumerate}
\item if $g > K$, then select an objective (\ref{eq:obj2})-minimizing non-basic column $j$ to enter basis, such that $g$ \textit{decreases} in the new basis $\cB'$.  
\item if $g < K$, then select an objective (\ref{eq:obj2})-minimizing non-basic column $j$ to enter basis, such that $g$ \textit{increases} in the new basis $\cB'$.  
\item if $g = K$, then select an objective (\ref{eq:obj2})-minimizing non-basic column $j$ to enter basis, such that $g$ \textit{remains the same} in the new basis $\cB'$. 
\end{enumerate}
When $g=K$, if the objective can no longer be decreased by introducing a non-basic column, then we terminate the pivoting procedure.

The key idea in the algorithm is the following.
By definition of a bfs, each solution computed via Simplex pivoting given constraints (\ref{eq:LPconstraints}) has $\z_{\bar{\cB}} = 0$, or $N$-sparse. 
Because we know which basic column would leave the basis $\cB$ corresponding to an entering non-basic column using (\ref{eq:pivot}), we can control the sparsity of the active sensors (columns in $\cB$ corresponding to entries in $\bEta$) by choosing carefully the entering non-basic column.
We can thus identify sparse feasible solutions while locally optimizing objective in (\ref{eq:obj2}).

\vspace*{-0.2cm}
\section{Numerical Results}
\label{sec:results}

\subsection{Simulation Settings}

We consider a 10m $\times$ 10m square area where $N=30$ wireless sensors are uniformly distributed and transmit their measurements in each time frame to the FC located at the centre. For the simulations, we essentially consider the same parameters as in~\cite{Ju14jan}. Namely, the path loss $L(d_n)$ where $d_n$ is the distance between the fusion center and sensor $S_n$ is assumed to follow $L(d_n)=L(d_0)+10\alpha \log_{10} (\frac{d_n}{d_0})$, with a 30dB average signal power attenuation at the reference distance $d_0=1$m. The pathloss exponent is fixed to $\alpha=2$~\cite{Kur17feb}. All channels are subject to block Rayleigh fading. Uplink-downlink channel reciprocity is assumed as in~\cite{Ju14jan}, namely $h_n=g_n$ for all sensors $S_n$. The noise power spectral density is assumed to be $N_0=-160$dBm/Hz, and the bandwidth to be 1MHz. 
The transmit power of the FC $P_e$ and maximum allowed transmit power for each sensor $P_{\mathrm{max}}$ are both set to $20$dBm. In (\ref{eq:xi}), we fix $\rho_n=0.9$ $\forall n$ and a normalized energy harvesting time $T_e=1$. We consider two sets of nodes, one with a high transmit probability of $p_n=0.9$ and the other with low transmit probability of $p_n=0.1$.

The correlated signal $\mathbf{x}$ is generated as follows: we consider the K-Nearest-Neighbor (KNN) graph of the WSN~\cite{Zhu12mar}, with $K=8$ and assigning edge weights $w_{n,k}=\exp^{-\frac{l_{n,k}}{\sigma^2}}$, where $l_{n,k}$ is the distance between nodes $n$ and $k$. Then, $\mathbf{x}$ is randomly generated following the Gaussian Markov Random Field with regard to this KNN graph, with mean vector $\mathbf{0}$ and precision matrix $\Q=L+\delta I$ where $\delta <<1$.

The proposed signal/power reconstruction method is compared to the following benchmark schemes.

\noindent 1. \underline{\emph{Reference, known power}}: the reference scheme in~\cite{Zhu12mar} based on graph-signal compressed sensing: in this scheme, a CS-based data reconstruction of the sensed data is performed by exploiting the sparse graph-signal representation in its Laplacian eigenbasis $\U=[\mathbf{u}_0,\mathbf{u}_1,\cdots, \mathbf{u}_{N-1}]$ or GFT matrix of the graph, where $\mathbf{u}_n$ is the eigenvector of the eigenvalue $\lambda_n$ of $\L$. The FC solves the optimization problem: $\min_x ||\mathbf{x}||_1$ s.t. $\mathbf{y}=\U_{\Omega}\H \mathrm{diag}(\bm{\eta}) \mathbf{x}$.
However, this scheme requires the knowledge of the active sensor nodes at the FC in order to extract $\U_{\Omega}$, as well as the power $\bm{\eta}$. Therefore, this scheme is evaluated under this ideal assumption.

\noindent 2. \underline{\emph{Reference, unknown power}}: the scheme in~\cite{Zhu12mar} explained above, but with unknown power $\bm{\eta}$ as in the proposed scheme. 

\noindent 2. \underline{\emph{Proposed Baseline}}: this scheme provides a baseline approach for solving this inverse separation problem by only considering the fidelity term and smoothness prior in (\ref{eq:opt}), i.e., without the sumlog probabilities of active sensors nor sparsity constraint on $\mathbf{\eta}$. 
Thus, the objective in Step 1 becomes $ \underset{\bEta}{\mathrm{min}} \| \mathbf{y}-\bPhi \H \mathrm{diag}(\hat{\x}_{i-1}) \bEta \|_2^2$ s.t. $\mathbf{0} \leq \bEta \leq \bEta_{\max}$, which can be solved using standard convex optimization tools. 

\noindent 3. \underline{\emph{Proposed, known power}}: the reconstruction only solves over $x$ while the transmit power $\bm{\eta}$ is perfectly known at the FC. This indicates the lower bound of the proposed scheme.

All schemes are evaluated in terms of Mean Square Error (MSE) performance for $\mathbf{x}$, defined as $\mathrm{MSE}=\frac{||\mathbf{x}-\mathbf{\hat{x}}||_2^2}{N}$, averaged over 1000 time frames.

\subsection{Simulation Results}

\begin{figure}[t]
\centering
\includegraphics[scale=0.34]{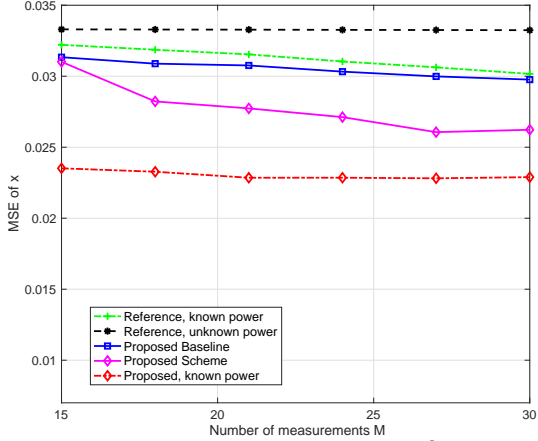} 
\vspace*{-0.6cm}
\caption{MSE performance of $\mathbf{x}$, $\sigma^2$=1}
\label{fig:MSE_sigma1}
\end{figure}

\begin{figure}[t]
\centering
\includegraphics[scale=0.34]{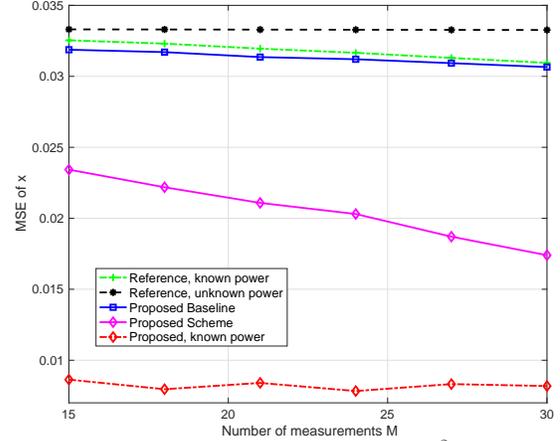} 
\vspace*{-0.6cm}
\caption{MSE performance of $\mathbf{x}$, $\sigma^2$=5}
\label{fig:MSE_sigma5}
\end{figure}

Figs. \ref{fig:MSE_sigma1} and \ref{fig:MSE_sigma5} show the MSE performance against varying numbers of observations $M$, for two different correlation parameters for $\mathbf{x}$, i.e., $\sigma^2=1$ for lower correlation and $\sigma^2=5$ for higher correlation, respectively. For simplicity, the average number of active sensors $K$ is here set to $M$, i.e., the number of observations. 
Fig. \ref{fig:MSE_sigma1} shows that even when the data correlation level is low, i.e., the unfavourable case for the proposed method, the \emph{Proposed Baseline} scheme achieves a lower MSE compared to the reference scheme of~\cite{Zhu12mar} that assumes perfect knowledge of power $\bm{\eta}$. Then, the \emph{Proposed Scheme} enables to further reduce the MSE performance.

Then, with a higher data correlation level, Fig. \ref{fig:MSE_sigma5} shows that the gain of our proposed scheme against reference schemes is further increased. 
By integrating the probabilities of active sensors and the sparsity constraint on $\mathbf{\eta}$ in the optimization, our proposed algorithm largely improves the performance achieved by the \emph{Proposed Baseline} scheme. 
Note that compared to Fig. \ref{fig:MSE_sigma1}, the lower bound of our proposed scheme, \emph{Proposed, known power}, is able to achieve a much lower  MSE in this case, since the smoothness prior on $\mathbf{x}$ is better suited when the underlying data exhibits higher correlations.  



\vspace*{-0.2cm}
\section{Conclusion}
\label{sec:conclude}
We consider an energy harvesting WSN, where the wireless sensors receive energy from the FC through downlink wireless power transfer and transmit their sensed data in the uplink phase. In this distributed setting, the variable transmit powers of sensor nodes are unknown at the FC. Hence, we formulate a joint signal / power restoration problem to recover sensed data given sparse observations from the few active sensors in each frame. 
Our proposed algorithm leverages on the smoothness of the data owing to spatial correlations, and the sparsity of the set of active sensors.  
Simulation results have shown that the proposed algorithm could achieve significantly low reconstruction error levels, even outperforming a CS-based benchmark scheme with perfect knowledge of the transmit power levels. By reducing the number of required observations for a given reconstruction level, the proposed method enables savings in spectral and energy resources. 


\addcontentsline{toc}{chapter}{References}
\bibliographystyle{IEEEtran}
\bibliography{IEEEabrv,ref2}





\end{document}